\documentclass[twocolumn]{aastex61}

\usepackage{xspace}
\usepackage{upgreek}
\usepackage{color}
\usepackage{natbib}
\usepackage{amsmath}
\usepackage{booktabs}
\usepackage{multirow}
\usepackage{tabularx} 
\usepackage{footnote}
\usepackage{lipsum}
\usepackage[normalem]{ulem}

\shorttitle{FRB repetition strategy}
\shortauthors{Connor \& Petroff}

\begin{document}
\title{On detecting repetition from fast radio bursts}

\author[0000-0002-7587-6352]{Liam Connor}
\affiliation{ASTRON, Netherlands Institute for Radio Astronomy, 
			Postbus 2, 7990 AA Dwingeloo, The Netherlands}
\affiliation{Anton Pannekoek Institute for Astronomy, 
			University of Amsterdam, Science Park 904, 
			1098 XH Amsterdam, The Netherlands}
\author{Emily Petroff}
\affiliation{ASTRON, Netherlands Institute for Radio Astronomy, 
			Postbus 2, 7990 AA Dwingeloo, The Netherlands}
\affiliation{Anton Pannekoek Institute for Astronomy, 
			University of Amsterdam, Science Park 904, 
			1098 XH Amsterdam, The Netherlands}

\email{liam.dean.connor@gmail.com}

\begin{abstract}
Fast radio bursts (FRBs) are bright, millisecond-duration 
radio pulses whose origins are 
unknown. To date, only one (FRB 121102) out of several 
dozen has been seen to repeat, though the extent to which it is 
exceptional remains unclear.
We discuss detecting repetition from 
FRBs, which will be very important for understanding 
their physical origin, and which also allows for host galaxy localisation. 
We show how the combination of 
instrument sensitivity, beamshapes, and individual FRB 
luminosity functions affect the detection of sources whose 
repetition is not necessarily described by a homogeneous Poisson process. 
We demonstrate that the Canadian Hydrogen Intensity Mapping Experiment (CHIME) 
could detect many new repeating FRBs for which host galaxies could be 
subsequently localised using other interferometers, 
but it will not be an ideal instrument for 
monitoring FRB 121102. If the luminosity distributions 
of repeating FRBs are given by power-laws 
with significantly more dim than bright bursts, CHIME's repetition 
discoveries could preferentially 
come not from its own discoveries, but from 
sources first detected with lower-sensitivity instruments like the 
Australian Square Kilometer Array Pathfinder (ASKAP) in fly's eye mode. 
We then discuss observing strategies for upcoming surveys, 
and advocate following up sources at approximately regular intervals
and with telescopes of higher sensitivity, 
when possible. Finally, we discuss doing pulsar-like 
periodicity searching on FRB follow-up data, 
based on the idea that while most pulses are undetectable, 
folding on an underlying rotation period could 
reveal the hidden signal. 

\end{abstract}
\section{Introduction}

Fast Radio Bursts (FRBs) are a new class of extragalactic
radio transient. 
They are characterized by their large
dispersion measures (DMs) (175--2600\,pc\,cm$^{-3}$) 
and short durations (30$\mu$s--30\,ms).
All FRB detections to date have been one-off events, except for FRB 121102. 
It was first detected with Arecibo \citep{spitler14}, but follow-up 
observations by \citet{spitler2016} found 10 
repeat bursts at the same DM as the initial detection. 
The repetition allowed for sub-arcsecond precision 
localization by the Very Large Array (VLA)
and the European VLBI Network (EVN) and provided the first direct 
host galaxy identification \citep{chatterjee2017, marcote2017, tendulkar2017}.
Eventually FRBs will be localized in real-time by interferometers
like Australian Square Kilometer Array Pathfinder \citep[ASKAP; ][]{bannister17} 
and \textit{realfast} \citep{law2018}, but no FRB has yet been 
successfully localized to a host galaxy through its discovery pulse.
Finding repeating FRBs and carrying out high time resolution 
VLBI follow-up is still a viable avenue for localizing host galaxies. 

Beyond allowing for localization, repetition also offers 
valuable clues about FRB progenitor 
models. Before the detection of FRB 121102's repetition, 
most theories postulated a cataclysmic origin
\citep{kashiyama2013, falcke2013, totani2013, zhang2014}. 
The local environment of FRB 121102 could play a role in 
the nature of its repetition. \citet{michilli2018} found that 
it resides in 
an extreme magneto-ionic environment that contributes a 
Faraday rotation measure (RM) of 
$\sim$\,$1.4\times10^5$\,rad\,m$^{-2}$.
Burst arrival times are clustered, individual bursts 
exhibit considerable time and frequency structure, 
and strong pulse-to-pulse brightness fluctuations are seen 
\citep{spitler2016,scholz2016,michilli2018}. 
While the sporadic repetition and 
pulse structure may be intrinsic, it is possible that 
they are coupled to the extreme plasma environment around 
the source. \citet{cordes2017} argued that FRB 121102 
might be behind a plasma lens that would magnify burst 
intensity and imprint time and frequency 
variations characteristic of caustics. 
Depending on the relationship between the 
resultant luminosity function of this amplification 
and the source's intrinsic luminosity function, magnification bias
may mean we preferentially see repeating FRBs that are associated with
extreme environments.

It remains unclear how many FRBs repeat and if their repetition 
is similar to FRB 121102. Of the more than 30 FRBs found so far\footnote{All 
publicly released FRBs are available on the FRB Catalogue; \url{www.frbcat.org}}
no others have been seen to repeat. Most of these bursts have 
had very little later follow-up, but some have 
been re-observed for tens and even hundreds of hours 
\citep{lorimer07, ravi2016, petroff-2015c}. However, meaningful 
constraints are difficult to achieve in the presence of clustered 
repetition \citep{connor-2016b,Oppermann18}.

In this paper we consider detecting repetition from 
FRBs with current and future surveys. In order to 
address questions of optimal observing strategy, we simulate 
repeating FRBs with arrival times that are not described by a
homogeneous Poisson process, and whose pulse intensities are
drawn from power-law luminosity functions. We consider the 
extent to which FRB 121102 may be unusual in the frequency 
and statistics of its repetition. 
We also discuss the possibility of repetition with an underlying period and 
show that for many brightness distributions, periodicity 
searches could offer higher signal-to-noise (S/N) 
than single-pulse searching.

\section{Repetition model}
\label{sec-formalism}

Before discussing the statistics of repetition it is 
important to distinguish the luminosity function 
of a single repeating source, which we will 
call $N(\mathcal{L})$, from both the global luminosity function 
of FRBs, $N_g(L)$, and their detected brightness distribution,
$N(\mathcal{F})$. The 
brightness distribution usually refers to the source counts 
in a given flux, fluence, or S/N bin (in the case of fluence this 
is $\log N$-$\log \mathcal{F}$). The $\log N$-$\log \mathcal{F}$ statistic 
encodes information about the 
volumetric distribution of FRBs, 
not just their intrinsic luminosities. It has important implications 
for survey design and the nature of the FRB population, and 
has been subject to considerable debate in the literature 
\citep{Oppermann16, vedantham2016, connor2017, macquart2018}. 
The ensemble luminosity function, $N_g(L)$,
with which we are not concerned here, refers to the number of 
bursts emitted with a certain luminosity across all FRBs; 
unlike $N(\mathcal{L})$ 
it would be meaningful even if all FRBs were one-off events. 

To describe the arrival times 
of repeat bursts, we follow the formalism of \citet{Oppermann18} and
use a generalization of Poissonian statistics known 
as the Weibull distribution. If repetition were 
well-described by a homogeneous Poisson process, 
then the probability of seeing $n$ repeats in some time 
interval $t$ given a rate $r$ would be,

\begin{equation}
p(n | t, r) = \frac{(rt)^n e^{-rt}}{n!}.
\end{equation}
\vspace{0.05 in}

\noindent This can also be written as an expected wait time, $\delta$,
between bursts. It will have the following exponential 
distribution, 

\begin{equation}
p(\delta | r) = r e^{-\delta r},
\end{equation}
\vspace{0.05 in}

\noindent where $r=\left<\delta\right>^{-1}$. 
This can be expanded to include a shape parameter,
$k$, which describes the clustering (or lack thereof) of events 
in time.
The probability density function (PDF) 
of time intervals for the Weibull distribution 
is,

\begin{equation}
\mathcal{W}(\delta | k, r) = \frac{k}{\delta} 
		\left [ \delta r \Gamma(1 + k^{-1}) \right ]^k 
		e^{-\left [\delta r \Gamma(1 + k^{-1})\right ]^k},
\label{eq-weibull}
\end{equation}
\vspace{0.05 in}

\noindent where $\Gamma(m)$ is the gamma function of 
some positive input, $m$. 
Eq.~\ref{eq-weibull} reduces to an exponential when $k$ 
is 1, so $k\neq1$ implies what we will call 
non-Poissonian statistics. 
For $k<1$ the probability of a new event \textit{decreases} 
with time, whereas $k>1$ indicates a monotonically \textit{increasing}
chance of an event. In other words, when the shape parameter is 
less than unity, events are clustered in time; when the shape 
parameter is large, event spacing tends toward uniformity. 

Interestingly, neutron stars span a large range of $(k, r)$ pairs. 
A typical radio pulsar is a good example of a $k$$>>$1 source, 
because as the shape parameter approaches infinity, the signal 
asymptotes to perfect periodicity. In that case, $1/r$ would simply 
be the spin period. 
The brightest giant pulses (GPs) from the Crab seem to follow Poissonian 
statistics, with $k\approx 1$ \citep{lundgren1995}. 
They are neither clustered in time, 
nor do they have predictable arrival times. If FRB 121102 does indeed 
come from a rotating neutron star, 
\citep{popov13,cordes2016,pen15,connor-2016a, spitler2016, lyutikov16} 
then this would be an example of a temporally clustered, $k<1$ source: 
\citet{Oppermann18} fit 
available data from FRB 121102, 
to a Weibull distribution and found best-fit parameters of 
$k$\,$=0.34^{+0.06}_{-0.05}$, $r$\,$=5.7^{+3.0}_{-2.0}$ 
repeats per day, strongly disfavouring 
homogeneous Poissonian repetition. They used data 
from 80 observations during which 17 pulses were 
found in a subset of just 7 pointings.

Repetition rates for an FRB must be treated in a similar way to 
FRB all-sky rates in general. It is not useful to give
an all-sky event rate of FRBs on its own. 
It must be accompanied by a complete brightness threshold,
for example $1.7^{+1.5}_{-0.9}\times 10^3$\,sky$^{-1}$\,day$^{-1}$ 
above 2\,Jy\,ms for widths between 0.13--32\,ms \citep{bhandari2018}.
Forecasting detection rates of surveys
with different sensitivities
requires further information about $\log N$-$\log \mathcal{F}$.
In the same way, repetition rates must be 
quoted not only with clustering parameters and an average interval, 
but also with a minimum brightness threshold. Again, forecasting expected repetition 
detection in different surveys will require knowledge about the particular 
source's $N(\mathcal{L})$. 
Thus the best fit values for 
FRB 121102 in \citet{Oppermann18} could be expanded to 
$k$\,$=0.34^{+0.06}_{-0.05}$, $r$\,$=5.7^{+3.0}_{-2.0}$ per day 
above 20\,mJy.

We parametrize the
luminosity function in the following way.

$$
\mathrm{d}N \propto \mathcal{L}^{-\gamma-1}\mathrm{d}\mathcal{L}
$$
\begin{equation}
N(>\mathcal{L}) \propto \mathcal{L}^{-\gamma}\,\,\,\,\,\,\,\,\gamma\neq 0.
\label{eq-lfunction}
\end{equation}
\vspace{0.05 in}

\noindent This equation gives the number of repeat 
bursts from the same FRB-emitting source that 
are greater than some luminosity, $\mathcal{L}$.
Our power-law assumption is partly motivated by 
the brightness distribution of giant pulses from 
Galactic pulsars, like the Crab \citep{cordes2004, zhuravlev2011, mickaliger2012, oronsaye15} 
or PSR B1937+21 \citep{Zhuravlev2013}, but also 
the distribution of burst energies and S/N from FRB 121102 
\citep{spitler2016, law2017}.

A boost in the observed repeat rate can be achieved 
during follow-up observation. 
For example, since the original detection was necessarily made 
in a blind search over a large number of DM trials, 
the phase space that must be 
searched by a dedispersion back-end in follow-up will 
shrink by a factor of $N_{\rm DM}$. This allows for a 
reduction in the S/N cutoff from the blind threshold, 
$\sigma_B$, to the targeted threshold, $\sigma_T$. 
Similarly, a boost can be achieved if the initial detection 
was found in an instrument's sidelobe, or if it 
was found with a lower-sensitivity telescope than 
the one searching for repetition. Assuming the 
initial detection was found with a 
system-equivalent flux density (SEFD) of $S_{\rm B}$
then followed up with $S_{\rm T}$, we can combine 
these with Eq.~\ref{eq-lfunction} to get 
the transformed rate,

\begin{equation}
r \rightarrow r \left( \frac{\sigma_B\, S_{\rm B}}{\sigma_T\, S_{\rm T}} 
 \sqrt{\frac{B_T}{B_B}}\right )^\gamma,
\label{eq-boost}
\end{equation}
\vspace{0.05 in}

\noindent where $B_B$ and $B_T$ are the bandwidths of 
the blind observation and the targeted one, respectively.

If either the individual FRB's brightness function is 
steep (i.e. $\gamma$ is large) or the ratio of 
SEFDs is large, the observed repeat rate of the 
source will change drastically. As an example, suppose 
$\gamma$ is 2, as with Crab GPs, and the source was 
initially detected in an instrument's sidelobe
but can be localized to less than a primary beam width.  
If the forward gain of the sidelobe 
is 15\,dB below that of on-axis, and a significance 
threshold was lowered from 10\,$\sigma$ to 5\,$\sigma$, 
then the detectable repeat rate of the FRB will increase 
by a factor of 4000---an example that may be significant 
for FRB 121102, as discussed in Sect.~\ref{sec-121102}. 
The same is true for follow-up 
observing with a more sensitive telescope. Assuming the same 
luminosity distribution, an FRB 
found with ASKAP in fly's eye mode will have an average 
detectable repeat rate that is $\sim$\,$10^4$ times larger 
when observed with Parkes, if the power-law continues over at 
least a couple of decades in $\mathcal{L}$. 

\section{Considerations for specific cases}
\subsection{FRB 121102}
\label{sec-121102}

The extent to which FRB 121102 is exceptional remains unclear. 
Some of its ostensibly unusual properties could, in principle, 
be explained by observational selection effects, and others 
no longer appear to be unique to that source.
For example, 
the non-power-law spectral behaviour found by \citet{spitler2016} 
and \citet{scholz2016}
seemed to set FRB 121102 apart from other FRBs, but 
recently more events have been discovered 
whose frequency structure is clumpy \citep{bannister17, Farah2018}.
The temporal structure 
that emerged after coherent dedispersion seemed different from other 
FRBs, many of which are unresolved \citep{ravi2017}. 
However, \citet{Farah2018} showed that FRB 170827, 
initially unresolved after being detected with incoherent dedispersion, 
also exhibits rich time and frequency structure when coherently 
dedispersed. 
The large rotation measure (RM) of FRB 121102, $\sim 10^5$ rad m$^{-2}$ 
\citep{michilli2018}, may also have been 
present in previous FRBs with polarization measurements, 
but intra-channel Faraday rotation would have smeared out 
their linear polarization \citep{petroff2015, keane2016}.

The most salient difference between FRB 121102 and other FRBs is 
that it is known to repeat. All other detected FRBs have been 
one-off events, despite, in some cases, considerable follow-up observations \citep{lorimer07,petroff-2015c,ravi2016}. 
However, as shown by \citet{connor-2016b}, when
repeat pulses are clustered in time  
strong constraints on repeat rate are difficult to achieve. 
Still, there are reasons to think FRB 121102 is particularly special 
in its repeatability. One is discovery bias, which refers to the 
fact that initial detections of phenomena tend to themselves 
be outliers. \citet{macquart2018} argued that the ultra-bright 
first FRB discovery, the Lorimer burst \citep{lorimer07}, should not be included 
in calculating population statistics for this reason. Indeed, 
there may already be evidence that FRB 121102 is exceptional 
in at least its frequency of repetition \citep{Palaniswamy2018}.
Another reason why FRB 121102 may have been found to repeat before others 
is that the initial detection \citep{spitler14} 
was in the Arecibo L-band Feed Array (ALFA) sidelobe 
where the forward gain was less than 1.7\,K\,Jy$^{-1}$
\citep{spitler2016}. With the seven-beam ALFA receiver, they 
used a grid of six pointings and found 10 repeat bursts 
\citet{spitler2016} all of which were in primary beams whose forward gains 
are between 8--10\,K\,Jy$^{-1}$. 
With Eq.~\ref{eq-boost}, 
the rate of FRB 121102 increased by a factor of 
approximately 8$^\gamma$ for the on-axis observing.

\begin{figure}
	\centering
          \includegraphics[clip, width=1.0\columnwidth]{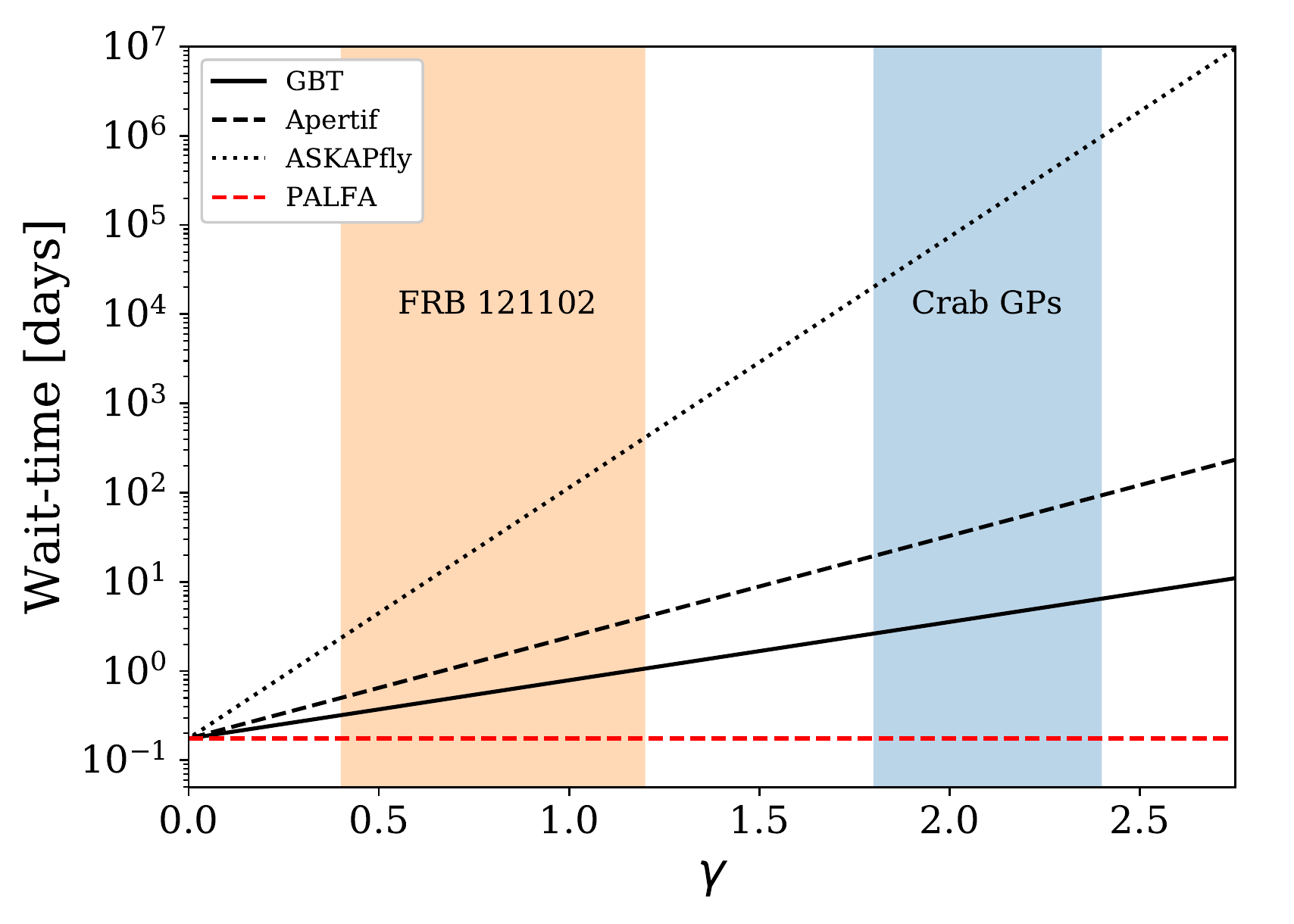}
	\caption{Average wait time between detectable FRB 121102 bursts 
			 for four different telescopes at 1.4\,GHz (The 
             Green Bank Telescope (GBT), Apertif, ASKAP in 
             fly's eye mode, 
             and the Arecibo ALFA receiver), 
             as a function of the cumulative
			 luminosity distribution power-law index $\gamma$. The 
			 horizontal line is the best-fit value of 5.7 events 
			 per day with ALFA found by \citet{Oppermann18}. This 
             figure shows the significance of brightness distribution 
             on effective repetition rate. In the case of FRB 121102, 
			 less sensitive instruments should not devote much 
             time to its monitoring, assuming the $\gamma$ values 
             found in \citet{law2017}.}
\label{fig-waittime}
\end{figure}

\citet{law2017} analyzed the repetition rate 
of FRB 121102 as a function of burst energetics. 
They find that its luminosity function is 
described by a power-law continuing over at least 
2.5 orders-of-magnitude, with $\gamma=0.7^{+0.5}_{-0.3}$.
In Fig.~\ref{fig-waittime} we show the expected wait time 
for a given telescope to see an FRB 121102 burst, 
plotted as a function of power-law index $\gamma$. 
We have highlighted the region of the 1--2\,GHz Crab GP's 
power-law index in blue, as well as the fit by 
\citet{law2017} in red. 
There is still significant uncertainty on the average 
repeat rate of FRB 121102, which would result in a 
re-scaling of the $y$-axis, but the purpose of the figure is to demonstrate 
the significance of $\gamma$ on repetition detection.

\subsection{CHIME}
The Canadian Hydrogen Intensity Mapping Experiment (CHIME)
is a transit instrument with no moving parts, observing 
between 400--800\,MHz
\citep{bandura2014, ng2017, chimbo2018}. It is expected to have 
the highest FRB detection rate of any upcoming survey due to its large 
collecting area and ability to search continuously 10$^3$ beams 
\citep{chawla2017, chimbo2018}. 
Even if the FRB rate were diminished at low frequencies due to 
scattering, free-free absorption, or spectral index, 
\citet{connor2016c} showed that using only the top quarter 
of CHIME's band between 700--800\,MHz---where the 
rate is known to be non-zero---would still result 
in 2--40 detections per day.

In Sect.~\ref{sec-advocacy} we prescribe observing strategies 
for detecting repeating FRB sources in the presence of non-Poissonian 
repetition and negative-power-law brightness distributions. 
For CHIME, which is not steerable and which records data continuously,
scheduling does not have many degrees of freedom.
Fortunately, transit instruments with North-South 
primary beams naturally apply the optimal 
observing strategy for clustered repetition, 
per unit time on source. With an East-West primary beam 
of 1--2$^\circ$, CHIME will have 0.3--0.6$\,\%$
of the sky visible at any given time. 
This means that once it has discovered a few hundred 
FRBs, CHIME will be doing repetition follow-up at almost 
all times because there will be a known source at most 
right ascensions. Therefore, even if the probability of 
seeing a single source repeat over the course of a year 
is low, CHIME will automatically follow up 
large numbers of FRBs and it will either produce a catalog of 
repeaters or establish the uniqueness of FRB 121102. 
If it does find repeaters, they can be followed up with 
longer-baseline interferometers like the VLA for localization.

One interesting implication of the effect of 
sensitivity on repetition rate is that 
CHIME may see more repeat pulses from FRBs originally detected at other, 
less sensitive instruments than from FRBs found with 
CHIME itself. Suppose each individual 
FRB repeats with a luminosity distribution 
$N(>\mathcal{L}) \propto \mathcal{L}^{-\gamma}$. 
Suppose also that CHIME, with SEFD $S_{\rm CH}$
has detected $N_{\rm CH}$ FRBs, which it can follow up daily. 
Then if some other survey X, with SEFD $S_{\rm X}$, has detected 
$N_{\rm X}$ bursts that are visible to CHIME, the 
condition, 

\begin{equation}
\left( \frac{S_{\rm X}}{S_{\rm CH}} \sqrt{\frac{B_{CH}}{B_X}}\right)^{\gamma} > 
\frac{N_{\rm CH}}{N_{\rm X}},
\label{equation-chimex}
\end{equation}
\vspace{0.05 in}

\noindent implies that CHIME will find more repeaters 
that were initially detected in survey X than detected at 
CHIME. 

As a simple example, if ASKAP in fly's eye mode 
were to find 50 bursts at 
declinations, $\delta$ $\gtrsim$\,$-10^\circ$
and CHIME amassed a set of 500 FRBs, we can 
calculate the values of $\gamma$ above which the relation 
in Eq.~\ref{equation-chimex} hold. Taking 
$S_{\rm CH} \approx 50\,\rm K / 1.3 \rm\,K\, Jy^{-1} \approx 38\,Jy$ 
and $S_{\rm ASKAP} = 1800\,\rm Jy$ \citep{bannister17}, 
then if $\gamma > 0.6$ CHIME will 
detect more repeaters from the ASKAP collection than from its own. 
This could also be true for Apertif \citep{leeu14} incoherent-mode detections.
And in general, if $\frac{N_{\rm CH}}{N_{\rm X}}$ is 
also the ratio of detection \textit{rates}, 
then the inequality will always be true.
Therefore, when possible, 
surveys like CHIME ought to search for repetition 
from known FRBs discovered by less sensitive instruments, 
potentially with a decreased S/N threshold at the initial 
detection's DM.

Though it sees the whole Northern sky each day,
a given source is only in CHIME's primary beam 
for $\sim$\,10 minutes, totaling roughly 60 hours 
per year. This fact, combined with the brightness function arguments in 
Sect.~\ref{sec-formalism} means it may not be 
a good tool for monitoring FRB 121102.
If FRB 121102's mean repeat rate with ALFA 
is 5.7 events per day, then one would expect roughly a 
dozen pulses per year if the source had a 
perfectly flat brightness distribution. 
But FRB 121102's 
luminosity function is not flat, and almost all of the 
bursts detected from FRB 121102 would not have been detected 
by CHIME, (or Apertif, or Parkes, or ASKAP). 
We have assumed that FRB 121102's repetition rate and 
brightness is the same at 
400--800\,MHz as it is at higher frequencies. 
The behaviour of FRB 121102 is not 
well contrained below 1\,GHz. Out of the 80 
observations in \citet{scholz2016}, 11 were 
taken with GBT's 820\,MHz receiver with 
200\,MHz bandwidth and no bursts were seen. 
However, given the highly non-Poissonian 
behaviour of FRB 121102 and the lower sensitivity 
at 820\,MHz due to less bandwidth, it is unsurprising 
that none was detected. 

\begin{figure}
	\centering
          \includegraphics[trim={1.5cm 0 1cm 0}, clip, width=1.05\columnwidth]{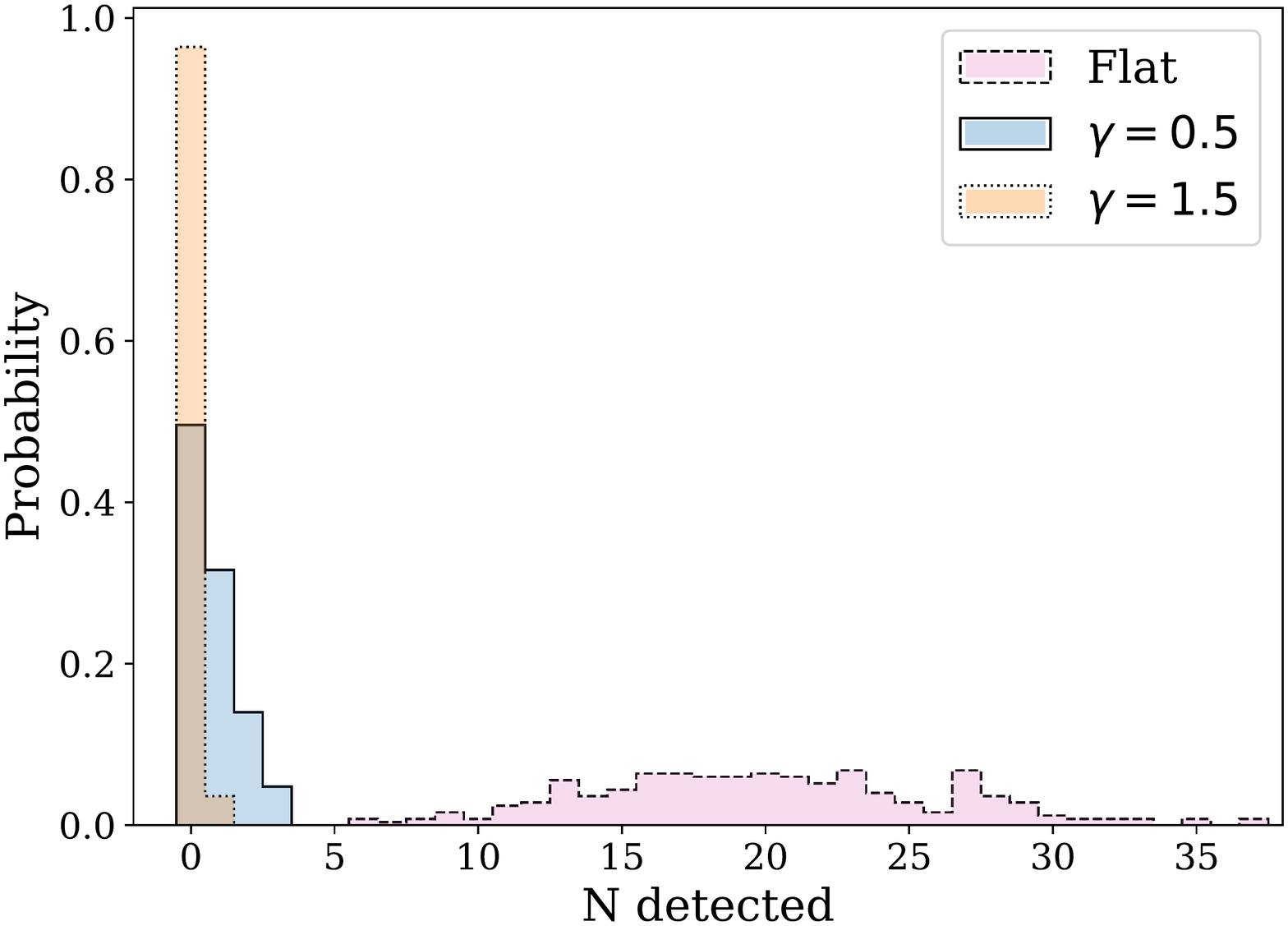}
	\caption{Distribution of the number of FRB 121102 burst detections in 
			 one year of observing  
			 with CHIME. We simulate events with 
			 the repetition statistics found 
			 by \citet{Oppermann18} and use three different pulse luminosity 
			 distributions. One (pink, dashed) does not account for 121102's 
             luminosity distribution and assumes Arecibo and CHIME 
             would see the same mean event rate. The other two 
             use power-laws such that 
             $N(>\mathcal{L}) \propto \mathcal{L}^{-0.5}$ (blue, solid), and
             $N(>\mathcal{L}) \propto \mathcal{L}^{-1.5}$ (orange, dotted).}
\label{fig-ch-121102}
\end{figure}

We have built a simple Monte Carlo simulation 
in which pulse arrival times are given by a 
Weibull distribution and each burst's energy
is randomly drawn from some input luminosity function.
In Fig.~\ref{fig-ch-121102} we show the 
number of expected FRB 121102 bursts detected per year 
with CHIME, assuming three different power-laws. 
This figure shows that unless FRB 121102's 
repeat rate turns out to be considerably larger than
the value used here, or $\gamma$ is smaller than $\sim$\,0.5, CHIME 
may see at most a few FRB 121102 bursts per year.

\section{Survey strategy}
\label{sec-advocacy}
We advocate doing follow up observations with 
more sensitive instruments than the detection survey, 
if possible. The Five-hundred-meter Aperture Spherical 
Radio Telescope (FAST) will have extraordinary sensitivity 
at 1.4\,GHz, with an expected forward gain of 18\,K\,Jy$^{-1}$ 
and system temperature of 20\,K \citep{Li2016}.
Therefore with an SEFD that is almost 2000 times lower 
than that of ASKAP in ``fly's eye'' mode, FAST would see 
an effective repeat rate that is $(2\times10^3)^\gamma$ times higher 
than ASKAP, assuming the power-law holds beyond those 
brightnesses. For single pixel instruments, the FoV mismatch between
large and smaller dishes would make this program difficult, 
because follow-up observations would have to be tiled. However, 
even in the extreme example of ASKAP fly's eye 
and FAST, the tiling problem is softened by their 
multi-pixel receivers. 
ASKAP's phased-array feed (PAF) allowed it to localize 
FRB 170107 to an 8$\times$8 arcminute 90$\%$ confidence region,
despite its nominal $\sim$\,1 degree beams \citep{bannister17}. 
FAST will initially have nineteen 2.9 arcminute beams in its 
1.4\,GHz multi-beam receiver, meaning the error region of 
FRB 170107 could be covered with between 1--7 pointings, depending 
on the angular separation of FAST's on-sky beams.

In the case of non-Poissonian repetition, the 
distribution of observing time has implications for 
detectability. If repetition is Poissonian, 
then detection statistics are affected only by 
total time on source. 
But if FRBs are significantly clustered in time then 
the worst observing strategy is to point at a source for a 
single long integration. 
\citet{connor-2016b} calculated this effect for $1/f$ noise, or a 
``pink distribution'' of arrival times in which bursts 
are significantly temporally correlated. 
\citet{Oppermann18} quantified it for 
a Weibull distribution with $k=0.35$
and $r=5.7$ per day, showing that
the chance of not seeing a burst during a 
single observation was four times larger than if observations 
were at approximately regular intervals and separated 
by roughly days, holding total time on source equal. 
The improvement becomes more extreme for lower values 
of $k$, i.e. higher clustering.

\begin{figure}
	\centering
          \includegraphics[trim={1.5cm .5cm 1cm .5cm}, clip, width=1.05\columnwidth]{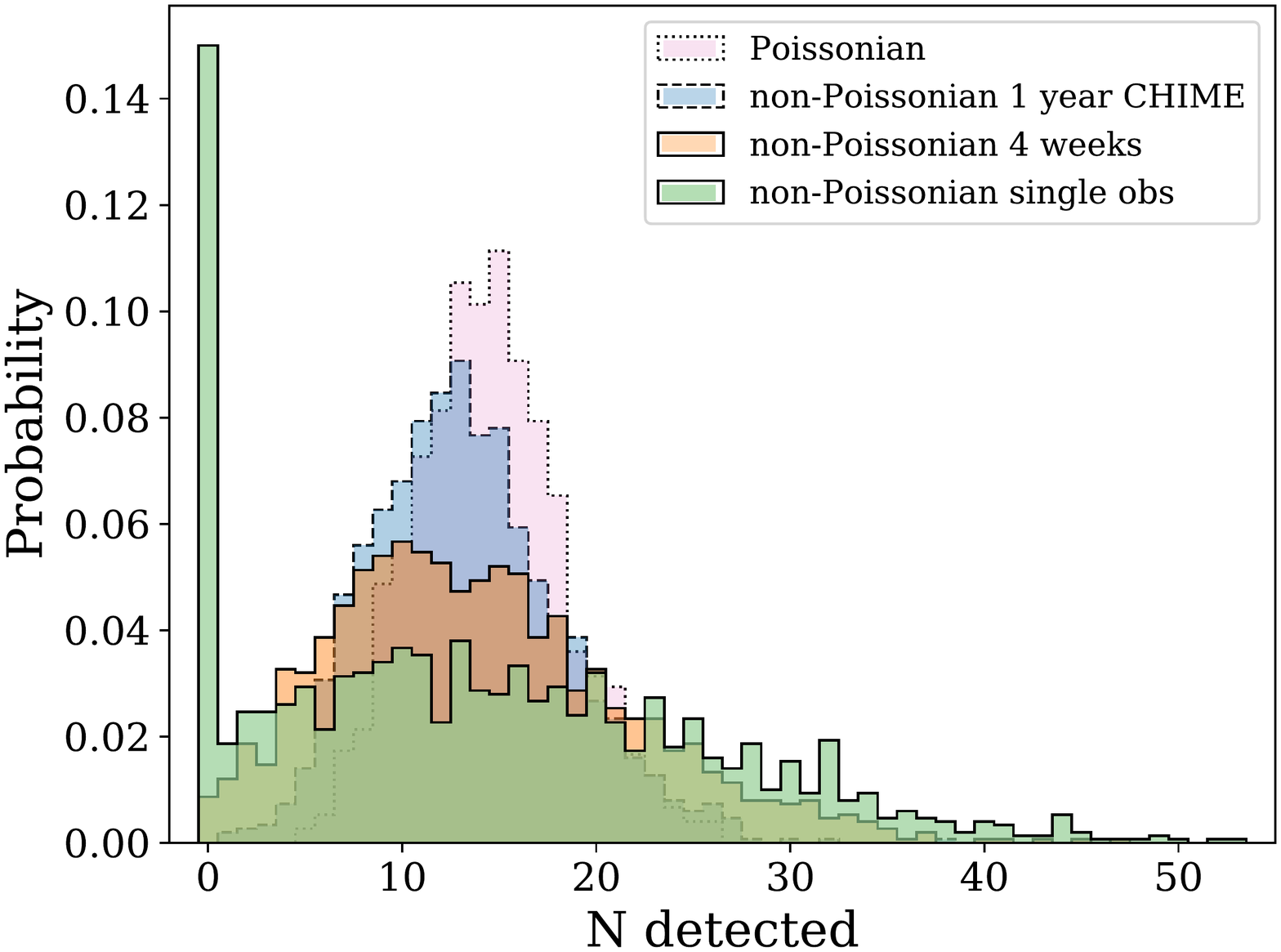}
	\caption{The distribution of number of 
    detected bursts from a repeating 
    source from a Monte Carlo simulation.
    We use four different follow-up strategies, 
    all of which total 60 hours on source, and 
    all with the same expected number of bursts.
	For the non-Poissonian bursts, we draw 
    repeat wait-times from a Weibull distribution 
    using the best-fit repetition statistics for 
    FRB 121102 found by \citet{Oppermann18}. 
    For a single 60-hour observation of a clustered
    repeater (green, solid), the probability of seeing zero events 
    is large despite an expected value of $\sim$\,$14$; if the time is spaced out with 
    daily 2-hour observations for four weeks 
    (blue, dashed), a clustered repeater's statistics 
    become near-Poissonian.}
\label{fig-strategy}
\end{figure}

As discussed, transit telescopes like CHIME automatically apply this  
periodic sampling function. However, for studies of a single 
FRB of particular interest, one may want more than 
ten minutes per source per day. For steerable surveys like 
Apertif, follow-up should be spread out over multiple 
observing sessions, switching between sources. 
For a clustered source, detection of a burst implies 
a higher probability of a subsequent burst. Therefore, 
telescopes that detect an FRB in real-time should stay on 
that source. 
In Fig.~\ref{fig-strategy} we show detection 
distributions from a simulated repeater. The results 
show that a single long observation of a clustered source 
can result in zero detected events, but regular short 
observations ($\lesssim2$\,hrs) reduces the clustered 
repeater to a near-Poissonian source. For a source 
whose repetition is not clustered in time, 
detectability is not affected by spreading observations 
out, lessening the risk of assuming all FRBs 
are like FRB 121102.

Priority of targets could be determined by 
brightness,
assuming there are more dim bursts than highly
energetic ones for a given repeater. It may also be useful to preferentially 
follow up recently discovered sources whose 
age lower-limit is the smallest, motivated by the idea 
that FRBs come from young neutron stars whose emitting 
window is limited \citep{connor-2016a, piro16, metzger17,
murase16}. 
Therefore, bright, recent sources like 
FRBs 170827 \citep{Farah2018}, 170107 \citep{bannister17}, and 180309\footnote{\url{http://www.astronomerstelegram.org/?read=11385}}
which were 60\,Jy, 
20\,Jy, and 100\,Jy, respectively, should be followed-up periodically 
by, e.g., Parkes. 

\subsection{Galactic FRBs}

An FRB from our own Galaxy would be extremely bright, 
and might be detected by a low-sensitivity, 
all-sky instrument, e.g. 
STARE\footnote{\url{www.caastro.org/files/64/3383921601/bochenek\_frb2018presentation\_upload.pdf}} or 
in other telescope's sidelobes \citep{tendulkar16}. This is 
because a Galactic FRB would be very rare, 
but would not require much forward 
gain to detect, so beam solid angle wins over sensitivity. It also 
means surveys that hope to preserve such signals should take 
measures not to throw them out as RFI, as they may show up 
as moderate-DM multi-beam detections.

Here we consider a Galactic FRB to be a short-duration radio pulse whose 
energetics are close to those of known extragalactic FRBs;
we do not include the Crab GPs that have been measured 
to date to be in this category. 
The probability of ever seeing such an 
event depends strongly on the statistics of FRB repetition. 
Suppose there are 5000 FRBs across the sky each day 
above some brightness threshold. 
If we imagine the extreme case, where all 5000 of those come 
from a single repeater, then every Galaxy in the 
observable Universe except one has zero FRB-emitting 
sources in it over the timescale of the typical emitting window. 
Even if the average repeat rate of 
FRBs is once per day---lower than FRB 121102---then there 
are only a few thousand galaxies in the Universe containing 
FRBs during an emitting window. 
Conversely, in the non-repeating case one simply has to wait 
long enough to see an FRB locally.
Therefore an observer in a random location sees either 
many ultra-bright Galactic FRBs or none. In reality 
there exists a continuum of FRB brightnesses, 
such that there are many more events below our surveys' 
detection thresholds than, in this example, 
a few thousand. This argument also depends on repeating 
sources being the dominant sub-class of FRBs.
Still, the statistics of repetition strongly 
affect the implied spatial distribution of burst-emitting sources. 

\subsection{Non-power-law $N(\mathcal{L})$}
FRB 121102 and giant pulses have power-law 
$N(\mathcal{L})$, but there is no guarantee that other repeating 
sources will have such a functional form;
brightness fluctuations in regular pulsars 
have been modelled with log-normal and  
Chi-squared distributions. 
However, even if the brightness function flattens 
out for very faint events, or decays exponentially,
follow-up with higher sensitivity 
will significantly improve detectability so long 
as there are more dim repeat bursts than bright ones. 

\subsection{FRB periodicity search}
Though no underlying periodicity has been found in 
FRB 121102's repetition, it is possible that other 
repeating sources will have a more easily detectable 
periodic signal. Therefore, it may be useful to do a 
periodicity search on one-off FRBs. This could be done 
either with an 
FFT-search or a fast folding algorithm (FFA)
to protect against missing long-period or low-duty cycle 
repeating sources \citep{staelin69,cameron17}.
If other sources have luminosity distributions 
like FRB 121102 in which there are more dim pulses than bright ones, 
de-dispersing to the known DM and folding on a range of rotation 
periods might pull the underlying signal out of the noise. 
The periodicity search could either be done during follow-up 
with a more sensitive instrument
or in data around the detection.
The latter may help for intermittent sources; the former 
would allow a hidden underlying signal to emerge more easily.
Periodicity in rotating radio transients (RRATs) was 
initially found using single-pulse time differencing \citep{mclaughlin2008},
but folding searches like the one we discuss here 
have been successfully used in sources such as 
RRAT J1819-1458 \citep{palliyaguru2011}.

The S/N optimality of periodicity searching vs. single-pulse 
detection depends primarily on the number of pulses in 
an observation, $N_p$, and the luminosity 
distribution of those pulses, $N(\mathcal{L})$ \citep{maclaughlin2003}.
Periodicity searches tend to achieve higher 
S/N if there are a large number of pulses in an observation, 
which is why they are often used for millisecond pulsars.
For a perfectly flat brightness distribution between some 
minimum brightness, $\mathcal{F}_{\rm min}$ and 
maximum $\mathcal{F}_{\rm max}$, 
\citet{maclaughlin2003} show that
FFT-periodicity searching is preferred 
over single-pulse search for $N_p\gtrsim 11$. For other 
power-law index, or indeed different functional forms like an exponential 
distribution, the values of $N_p$ for which periodicity searching 
results in higher S/N may be larger. 

It may also be that repeating FRBs 
do not have any underlying periodicity. 
If this is the case, an excess of events 
that are below the single-pulse threshold 
but above, say, 3\,$\sigma$ could be looked for in the dedispersed 
time-stream at the FRB detection DM. The cumulative S/N distribution
of all samples (binned to the width of the initial burst)
could be compared to the same data dedispersed to other DMs 
in order to look for statistically significant differences in their tails. 
While there is no guarantee that these techniques would 
uncover repetition,
periodicity searching a small range of DMs or 
making a histogram of S/N is relatively low cost, 
whereas demonstrating repetition with a folded 
spectrum would be highly informative.

\section{Conclusions}

In this letter we have investigated the interplay between 
instrumental effects, survey strategy, and the detectability 
of FRB repetition from sources whose burst arrival times 
are not necessarily described by a homogeneous Poisson process. 
We summarize our findings as follows:

\begin{itemize}

\item If other repeating FRBs have luminosity functions 
similar to FRB 121102 or the Crab's giant pulses, in that 
there are many more dim bursts than bright ones, observed 
repeat rates can be greatly boosted by doing follow-up observations with 
higher sensitivity instruments than the original detection telescope. 
The same is true for sidelobe detections that are followed up on-axis, 
which is how FRB 121102 was found to repeat, and may have contributed 
to it being the first repeater discovered.

\item North-South transit telescopes like CHIME are well-suited 
to search for repetition in FRBs whose bursts are clustered in time,
allowing for arcsecond localization by longer-baseline interferometers.
However, CHIME will not be an ideal instrument for monitoring the 
one known repeater, FRB 121102.
\vspace{0.1in}
\item Repetition found by CHIME may preferentially come from FRBs initially 
detected at other, lower-sensitivity instruments like ASKAP or 
Apertif incoherent mode.

\item The probability of ever detecting an ultra-bright 
Galactic FRB depends strongly on the statistics of repetition.

\item Periodicity searches at the known DM of one-off FRBs could 
reveal repetition in their folded spectra, ideally 
carried out with more sensitive instruments than the detection telescope. 
Such searches could also be 
done in pre-existing data around the initial detection.  
\end{itemize}

\begin{acknowledgements}

We thank Jason Hessels for helpful notes 
on the manuscript. We also thank our referee for 
offering useful feedback.
LC and EP receive funding from the European Research 
Council under the European Union's Seventh Framework Programme (FP/2007-2013) / ERC Grant Agreement n. 617199.

\end{acknowledgements}

\bibliography{repeat_paper}
\bibliographystyle{aastex61}

\end{document}